\documentclass{article}

\usepackage[nonatbib, final]{neurips_2024}

\usepackage[utf8]{inputenc} 
\usepackage[T1]{fontenc}    
\usepackage{hyperref}       
\usepackage{url}            
\usepackage{booktabs}       
\usepackage{amsfonts}       
\usepackage{nicefrac}       
\usepackage{microtype}      
\usepackage{xcolor}         
\usepackage{natbib}
\usepackage{wrapfig}

\usepackage[russian,english]{babel}

\usepackage{ tipa }

\newcommand{\textcyr}[1]{\begingroup\fontencoding{T2A}\selectfont #1\endgroup}

\usepackage{tabularx} 
\usepackage{graphicx} 
\usepackage{booktabs} 

\usepackage{makecell} 
\usepackage{array} 
\usepackage[T1]{fontenc}

\usepackage[utf8]{inputenc}

\usepackage{microtype}

\usepackage{inconsolata}
\usepackage{amsmath}
\usepackage{amsfonts}%
\usepackage{amssymb}%
\usepackage{graphicx}
\usepackage{url}
\usepackage{booktabs}
\usepackage{multirow}
\usepackage{placeins}
\usepackage{comment}
\usepackage{multirow}
\usepackage{url}
\usepackage{subcaption}
\usepackage{multirow} 

\usepackage{xcolor}

\definecolor{green}{rgb}{0.0, 0.5, 0.0}
\definecolor{purple}{cmyk}{0.5, 1.0, 0.0, 0.2}
\usepackage{graphicx}

\usepackage{booktabs}
\usepackage{longtable}

\usepackage[normalem]{ulem}
\definecolor{insertioncolor}{RGB}{0, 0, 255}
\definecolor{substitutioncolor}{RGB}{255, 165, 0}
\definecolor{deletioncolor}{RGB}{255, 0, 0}

\newcommand{\insertion}[1]{\textcolor{insertioncolor}{\underline{#1}}} 
\newcommand{\substitution}[1]{\textcolor{substitutioncolor}{#1}} 
\newcommand{\deletion}[1]{\textcolor{deletioncolor}{\sout{#1}}} 

\usepackage[utf8]{inputenc}
\usepackage[russian,english]{babel}

\usepackage{ tipa }

\title{CA-SSLR: Condition-Aware Self-Supervised Learning Representation for Generalized Speech Processing}

\author{%
  Yen-Ju Lu$^\dagger$, Jing Liu, Thomas Thebaud$^\dagger$, Laureano Moro-Velazquez$^\dagger$, \\ 
  \textbf{Ariya Rastrow}, \textbf{Najim Dehak$^\dagger$}, \textbf{Jesus Villalba$^\dagger$} \\
  $^\dagger$Center for Language and Speech Processing, Johns Hopkins University \\
  \texttt{\{ylu125, tthebau1, laureano, ndehak3, jvillal7\}@jhu.edu} \\
}

\begin{document}

\maketitle

\begin{abstract}
We introduce Condition-Aware Self-Supervised Learning Representation (CA-SSLR), a generalist conditioning model broadly applicable to various speech-processing tasks. 
Compared to standard fine-tuning methods that optimize for downstream models, CA-SSLR integrates language and speaker embeddings from earlier layers, making the SSL model aware of the current language and speaker context.
This approach reduces the reliance on the input audio features while preserving the integrity of the base SSLR. 
CA-SSLR improves the model’s capabilities and demonstrates its generality on unseen tasks with minimal task-specific tuning.
Our method employs linear modulation to dynamically adjust internal representations, enabling fine-grained adaptability without significantly altering the original model behavior.
Experiments show that CA-SSLR reduces the number of trainable parameters, mitigates overfitting, and excels in under-resourced and unseen tasks.
Specifically, CA-SSLR achieves a 10\% relative reduction in LID errors, a 37\% improvement in ASR CER on the ML-SUPERB benchmark, and a 27\% decrease in SV EER on VoxCeleb-1, demonstrating its effectiveness.
\end{abstract}


\section{Introduction}
\label{sec:intro}
The emergence of Self-Supervised Learning Representations (SSLRs) models has revolutionized speech processing, setting new standards in the field. Pioneering models like Wav2vec 2.0~\cite{baevski2020wav2vec}, HuBERT~\citep{hsu2021hubert}, and WavLM~\citep{chen2022wavlm} leverage unlabeled audio data to learn rich representations of spoken language. These models are pivotal in a wide range of applications, including Speech Recognition (ASR)~\citep{chang2021exploration}, Speaker Verification (SV)~\citep{chen2022does,tak2022automatic}, Language Identification (LID)~\citep{bartley2023accidental}, and Speech Translation (ST)~\citep{tang2022unified}. Benchmarks such as SUPERB~\citep{yang2021superb} and ML-SUPERB~\citep{shi2023ml} have been crucial in evaluating SSL model performance, providing standardized tasks.

Although SSLR training approaches combine speech from various sources, these models learn representations solely from unpaired audio-only data. When extending SSLR features to multilingual scenarios and low-resource languages, unsupervised training limits the model's ability to distinguish between different languages, resulting in unified features for all languages. Additionally, labeling all SSL training data with language and speaker information requires significant human effort and is impractical. Thus, a post-training conditioning approach is more favorable. In other fields, methods like~\citep{zhang2023adding} and IP-Adaptor ~\citep{ye2023ip} in image processing, and CTRL~\citep{keskar2019ctrl} in NLP, have successfully integrated conditioning into pretrained models, demonstrating potential applications for speech processing.

In response to these challenges, we propose Condition-Aware Self-Supervised Learning Representation (CA-SSLR), a generalist conditioning model applicable to various speech-processing tasks such as language identification, multilingual speech recognition, and speaker verification. Unlike standard adaptation methods that optimize the SSLR parameters for downstream models, CA-SSLR integrates language and speaker embeddings from earlier layers, making the SSLR aware of the current language and speaker context. This technique enables the creation of models that perform multiple tasks with a single adapted SSL encoder by strategically injecting conditional adapters into each encoder block while freezing the pretrained encoder weights. CA-SSLR follows a hierarchical self-adaptation structure, where adapters at each layer are conditioned on intermediate task-specific embeddings estimated from lower layers. Attention mechanisms and linear modulation dynamically adjust scaling and biasing, tailoring the model’s response at each time step. The initialization techniques allow the conditioning module to perform identity transformations, ensuring the existing model behavior is maintained when incorporating new conditions. This approach reduces the number of trainable parameters, mitigates overfitting, and avoids catastrophic forgetting. We conduct experiments on three popular types of multilingual speech processing tasks—ASR, LID, and SV to demonstrate the versatility and efficiency of CA-SSLR.

This work's main contribution is introducing a novel method for conditioning the SSLRs with limited supervised labels. This leads to generalized speech representation with improved performance using minimal trainable parameters and maintains the model's behavior. This includes:
\vspace{-2mm}
\begin{itemize}
\itemsep0em 
\item \textbf{Hierarchical Dynamic Conditioning}: We design attention-based conditional adapters and integrate them into the SSL model. Our approach dynamically tailors the model's behavior to the input language and speaker characteristics at each time step, which are periodically estimated from previous layers.

\item \textbf{Preservation of Pre-trained Weights with Efficient Parameter Utilization}: The model capitalizes on the knowledge of the foundational model pre-trained weights and introduces lightweight adapters that modulate the encoder hidden representation by a scalar $\gamma$ and bias $\beta$, significantly reducing the trainable parameters. This strategy ensures more stable and parameter-efficient training.

\item \textbf{Harmonized Task Compatibility with Notable Performance Improvements}: Our experiments show that CA-SSLR reduces the number of trainable parameters, mitigates overfitting, and excels in under-resourced and unseen tasks. Specifically, CA-SSLR achieves an 27\% relative reduction in LID errors, a 37\% improvement in ASR CER on the ML-SUPERB benchmark, and a 27\% decrease in SV EER on VoxCeleb-1. These results highlight CA-SSLR’s effectiveness in enhancing multilingual SSLRs while also lowering computational costs for multitask fine-tuning.

\end{itemize}

\section{Related Work}
\paragraph{Self-supervised learning representation.}
\label{sec:ssl}
Self-Supervised Learning (SSL) models, epitomized by Wav2Vec 2.0~\citep{baevski2020wav2vec}, HuBERT~\citep{hsu2021hubert}, and WavLM~\citep{chen2022wavlm}, have significantly advanced speech processing by leveraging vast amounts of unlabeled audio data. These models excel in extracting 
rich speech representations, capturing its intricate acoustic, phonetic, and semantic nuances. 
These models are fine-tuned on smaller, labeled datasets to adapt the generic representations for specific tasks, achieving impressive results. 

In the realm of cross-lingual speech representation, Wav2Vec 2.0-XLSR (Cross-Lingual Speech Representation)~\citep{babu2021xls} takes a significant leap forward. It builds on the robust architecture of Wav2Vec 2.0 but is pre-trained on a diverse, multilingual dataset, learning universal representations transferable across languages.
Similarly, mHuBERT (Multilingual HuBERT)~\citep{lee2021textless} extends the foundational HuBERT model to process multiple languages effectively. 
This makes them immensely powerful for multilingual speech recognition and understanding tasks. The benchmark for the SSL models also extends from monolingual SUPERB \citep{yang2021superb} to multilingual ML-SUPERB \citep{shi2023ml}, building new standards for the SSLR models.

\label{sec:cond_mod}

\paragraph{Adaptation Methods.}
In many studies~\cite{yang2021superb,shi2023ml,chen2023improving}, the SSLR remains frozen while decoders are trained for a specific task. Since the encoder is shared across all tasks, this approach offers the advantage that it allows us to evaluate multiple tasks on a given speech signal with just one encoder run. However, systems of this kind often exhibit poorer performance when compared to those incorporating some degree of adaptation of the SSLR to the target task. The latter can involve fine-tuning the entire SSL encoder~\cite{chen2022wavlm}, a subset of its layers, or introducing lightweight adapters~\citep{chen2023exploring} within its layers. Unfortunately, this results in employing a distinct encoder per task, leading to a large increase in computational load that scales linearly with the number of tasks to be assessed.  

\paragraph{Conditioning Pre-trained Models.}
Image processing has successfully integrated conditioning into pretrained models using methods like ControlNet ~\citep{zhang2023adding} and IP-Adaptor ~\citep{ye2023ip}. ControlNet allows for precise control over generated images by incorporating additional conditions such as edge maps or sketches, while IP-Adaptor uses small-scale adapter modules to adjust the model's behavior based on specific conditions without altering the pre-trained model's parameters. These techniques have achieved significant success and offer insights for potential applications in speech processing. Similarly, in Natural Language Processing (NLP), models like the Conditional Transformer Language Model (CTRL) ~\citep{keskar2019ctrl} have introduced conditioning to improve model performance. CTRL uses control codes to guide text generation based on specified attributes like style or domain, allowing for efficient adaptation without extensive retraining. The successes in image processing and NLP highlight the potential for conditioning pre-trained SSLRs in speech processing. 

\paragraph{Hierarchical Conditioning.}
Hierarchical models have been used in previous speech models. \citep{sanabria2018hierarchical} proposes a multi-task ASR model that improves intermediate representations by performing Connectionist Temporal Classification at different levels of the network with targets of different granularity. Essentially, representations in lower layers are used to predict character tokens, while higher layers predict subword units with growing vocabulary sizes--from 300 to 10k subword units in the last layer. \citep{chen2023improving} further explored this by integrating hierarchical conditional layers within the ASR decoder, using ASR tokens predicted from preceding layers to inform subsequent layers.

\begin{figure}[t]
    \centering
    \begin{subfigure}[t]{0.45\columnwidth}
        \centering
        \includegraphics[width=\linewidth]
        {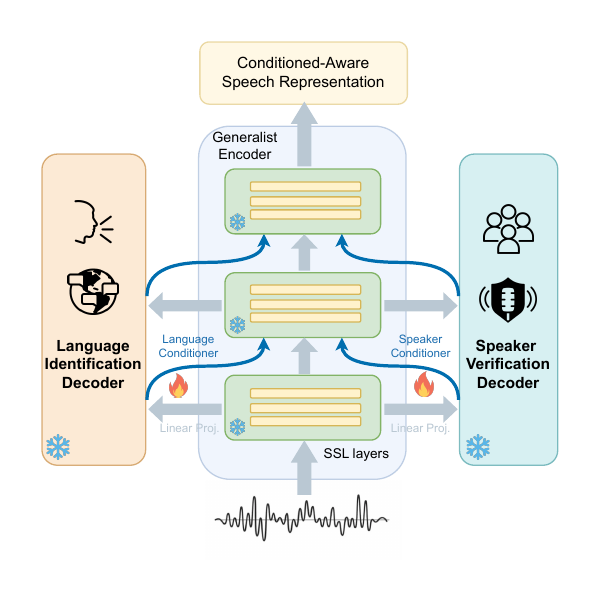}
        \caption{CA-SSLR improves SSL features by integrating intermediate LID/SV conditions, keeping pre-trained parameters frozen.}
        \label{fig:salli}
    \end{subfigure}
    \hfill
    \begin{subfigure}[t]{0.52\columnwidth}
        \centering
        \includegraphics[width=\linewidth]
        {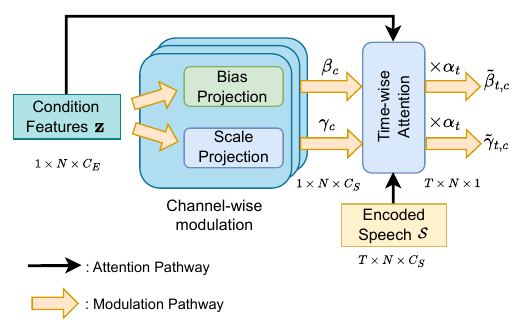}
        \caption{The trainable time-channel attention conditioner for integrating language and speaker prediction in CA-SSLR. It predicts bias $\tilde{\beta}$ and scale $\tilde{\gamma}$ using condition feature $\mathbf{z}$.}
        \label{fig:lli}
    \end{subfigure}
    \caption{CA-SSLR scheme and its time-channel attention conditioner. Only the conditioner and linear projections for the decoders are trainable, and all other parameters are frozen during adaptation.}
    \label{fig:combined}
    \vspace{-0.3cm}
\end{figure}





\section{Methodology} 
\label{sec:model}


\begin{figure}[t]
    \centering
    \begin{subfigure}[t]{0.76\columnwidth}
        \centering
        \includegraphics[width=\linewidth]
        {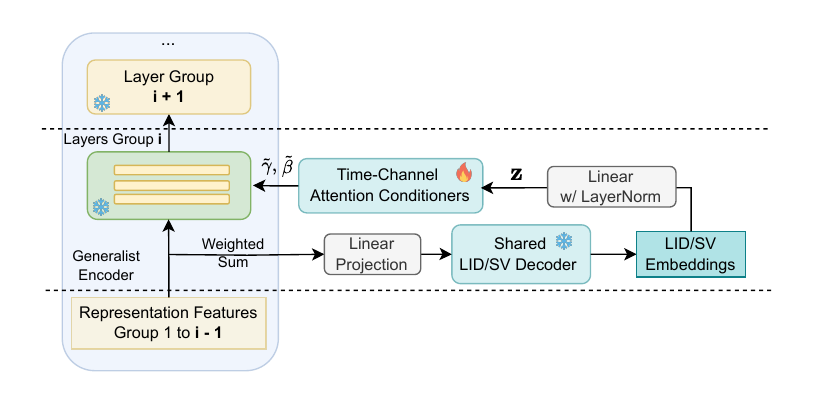}
        \caption{Hierarchical conditioning with TCACs to generate feature \( \mathbf{z} \) and modulate layers with scale \( \tilde{\gamma} \) and bias \( \tilde{\beta} \).}  
        \label{fig:ca-sslr-condition} 
    \end{subfigure}
    \hfill
    \begin{subfigure}[t]{0.23\columnwidth}
        \centering
        \includegraphics[width=\linewidth]
        {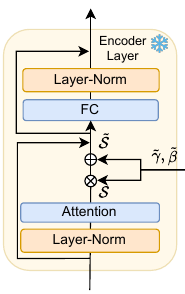}
        \caption{SSLR layer with conditioning integration, transforming \( \mathbf{S} \) into \( \tilde{\mathbf{S}} \).}
        \label{fig:ca-layer}
    \end{subfigure}
    \caption{Architecture of the CA-SSLR model employing hierarchical self-conditioning with Time-Channel Attention Conditioners (TCACs).}
    \label{fig:ca-hier-cond}
    \vspace{-0.3cm}
\end{figure}

We propose Condition-Aware SSLR (CA-SSLR), designed to serve as a universal encoder for multiple downstream speech tasks. CA-SSLR enhances pre-trained SSL models by integrating intermediate LID and SV predictions to condition and adapt subsequent layers dynamically. This approach allows the model to capture essential language and speaker characteristics, refining its outputs progressively and making it particularly effective in multilingual and multispeaker scenarios.

Figure~\ref{fig:salli} illustrates the overall architecture of CA-SSLR. The model consists of a frozen SSL encoder augmented with trainable conditioners and lightweight task-specific decoders. The conditioner modulates the hidden representations of the SSL encoder layers based on conditioning features derived from intermediate LID and SV embeddings. This hierarchical conditioning mechanism enables the model to adapt dynamically to different input conditions while keeping the majority of the pre-trained parameters fixed.
In the following sections, we detail the components of CA-SSLR, starting with the conditioner module and then explaining how it integrates into the overall architecture. We also describe the incremental training strategy employed to incorporate conditioning information without catastrophic forgetting.

\subsection{Channel-wise and Time-wise Attention Conditioner}
\label{sec:llc}

A central component of CA-SSLR is the channel-wise conditioner (CC) or the time-channel attention conditioners (TCAC), which modulates the SSL encoder's hidden representations based on conditioning features.
As depicted in Fig.~\ref{fig:lli}, the TCAC tasks the latent representations $\mathbf{S}^{(l)}\in\mathbb{R}^{C\times T}$ from layer $l$ of the SSL encoder and a conditioning feature vector $\mathbf{z}\in\mathbb{R}^R$, derived from intermediate LID or SV embeddings.
The TCAC outputs modulated latent representations $\mathbf{\tilde{S}}^{(l)}$ by applying time-channel-dependent scaling and bias:
\begin{align}
\label{eq:tca}
\mathbf{\tilde{S}}_{t,c}^{(l)}
=\text{TCAC}(\mathbf{S}_{t,c}^{(l)},\mathbf{z}) 
&= \tilde{\gamma}_{t,c}^{(l)}(\mathbf{z}, \mathbf{S}^{(l)})\mathbf{S}_{t,c}^{(l)} + \tilde{\beta}_{t,c}^{(l)}(\mathbf{z}, \mathbf{S}^{(l)}) 
\end{align}
where $t$ and $c$ index time and channel dimensions, respectively. The scales $\tilde{\gamma}_{t,c}^{(l)}$ and biases $\tilde{\beta}_{t,c}^{(l)}$ are products of time-dependent and channel-dependent components:
\begin{align}
\tilde{\gamma}_{t,c}^{(l)}(\mathbf{z}, \mathbf{S}^{(l)})=\alpha_{t}^{(l)}(\mathbf{z},\mathbf{S}^{(l)}) \times \gamma_{c}^{(l)}(\mathbf{z})  \qquad
\tilde{\beta}_{t,c}^{(l)}(\mathbf{z}, \mathbf{S}^{(l)})=\alpha_{t}^{(l)} (\mathbf{z},\mathbf{S}^{(l)})\times \beta_{c}^{(l)}(\mathbf{z})
\end{align}
The channel-dependent scales $\gamma^{(l)} \in \mathbb{R}^C$ and biases $\beta^{(l)} \in \mathbb{R}^C$ are computed via linear transformations of the conditioning feature, similar to feature-wise linear modulation~\citep{perez2018film}:
\begin{align}
   &\gamma^{(l)}(\mathbf{z})= \mathbf{W}_\gamma^{(l)} \mathbf{z} + \mathbf{b}_\gamma^{(l)} \quad 
   \beta^{(l)}(\mathbf{z})= \mathbf{W}_\beta^{(l)} \mathbf{z} + \mathbf{b}_\beta^{(l)} \quad \mathrm{with} \;.
\end{align}
The time-dependent scales $\alpha^{(l)}\in \mathbb{R}^T$ are obtained with an additive attention mechanism as
\begin{align}
\alpha_t^{(l)}(\mathbf{z},\mathbf{S}^{(l)})= \mathbf{v}_\alpha^\mathrm{T} f(
 \mathbf{W}_\alpha^{(l)}
 \begin{bmatrix}
 \mathbf{S}_t^{(l)} \\
 \mathbf{z}\\
 \end{bmatrix}
 +\mathbf{b}_\alpha^{(l)} )
\end{align}
where $f(.)$ is a ReLU non-linearity, $\mathbf{W}_\alpha^{(l)}\in \mathbb{R}^{C'\times (C+R)}$, $\mathbf{b}_\alpha^{(l)}\in\mathbb{R}^{C'}$, and $\mathbf{v_\alpha}\in\mathbb{R}^{C'}$.
The conditioning feature $\mathbf{z}$ is obtained by processing the intermediate embeddings $\mathbf{e} \in \mathbb{R}^E$ from the LID or SV decoders, as $\mathbf{z} = \mathrm{LayerNorm}(\mathbf{W} \mathbf{e} + \mathbf{b})$, where $\mathbf{W} \in \mathbb{R}^{R \times E}$ and $\mathbf{b} \in \mathbb{R}^R$ are shared linear transformation parameters, and $\text{LayerNorm}(\cdot)$ denotes layer normalization.
%
In scenarios where time-based modulation is unnecessary, the model can switch to the simpler Channel-wise Conditioner (CC) by using only the channel-dependent components $\gamma$ and $\beta$.
This flexibility in conditioning design enables the model to be tailored to various speech tasks with differing complexity requirements.
By integrating these conditioning methods, CA-SSLR dynamically adapts its internal representations based on language and speaker characteristics. This mechanism enables the integration of conditioning into the model's latent representations without altering the pre-trained encoder's original parameters.


\subsection{Hierarchical Self-Conditioning in CA-SSLR}
\label{sec:hierarchical}
Building upon the TCAC module, CA-SSLR employs a hierarchical self-conditioning mechanism within the SSL encoder layers. As shown in Figure~\ref{fig:ca-hier-cond}, the SSL encoder is partitioned into layer groups, with TCACs inserted after the attention module in each layer to modulate hidden representations based on updated conditioning features.
The model aggregates SSL features through a weighted sum, combining outputs from all preceding layer groups. These aggregated features are then provided to the LID and SV decoders, where LID and SV embeddings are extracted and processed through a linear layer followed by layer normalization to create the conditioning feature $\mathbf{z}$ for the TCACs.

 
The conditioning feature $\mathbf{z}$ is re-estimated at intervals—every three layers for LID and every six layers for SV—using the aggregated SSL features from previous groups. This hierarchical design progressively refines the model's representations, adapting to the input's language and speaker characteristics at different depths of the network. For example, the initial SSL layer group captures basic language and speaker characteristics, generating embeddings that condition the next group of layers via TCACs. This ongoing refinement allows the model to dynamically adapt based on intermediate predictions, resulting in a context-aware and dynamic representation.

Each layer group uses distinct TCAC parameters, enabling tailored scaling and bias adjustments at different stages of the model. Notably, only the TCACs and the linear projections for the decoders are trainable, while all other SSL encoder parameters remain fixed during the conditioning insertion. This design minimizes overfitting and accelerates training due to the smaller number of trainable parameters. 
This hierarchical self-conditioning mechanism enables the model to dynamically capture diverse aspects of input audio, making it a robust tool for comprehensive speech analysis.



\subsection{Incremental Training Strategy}
\label{sec:incremental}

Incorporating new components into a pre-trained SSL encoder poses the risk of catastrophic forgetting, where the model loses previously acquired knowledge. To mitigate this, we adopt an incremental training strategy that gradually integrates the conditioning information.
We initialize the TCAC parameters to ensure that the initial modulated features are identical to the original SSL features. Specifically, we set the initial values such that $\alpha_t = 1$ for all $t$, $\gamma_c = 1$, and $\beta_c = 0$ for all $c$. According to Eq.~\eqref{eq:tca}, this initialization means that $\mathbf{\tilde{S}}_{t,c}^{(l)} = \mathbf{S}_{t,c}^{(l)}$ at the start of training, allowing a smooth transition from the pre-trained model to the conditioned model.

When multiple conditioning features are involved, such as both LID and SV, we compute separate scaling and bias parameters for each:
\begin{equation}
\alpha_{\text{total}} = \alpha_{\text{LID}} \times \alpha_{\text{SV}}, \quad  \gamma_{\text{total}} = \gamma_{\text{LID}} \times \gamma_{\text{SV}}, \quad \beta_{\text{total}} = \beta_{\text{LID}} + \beta_{\text{SV}}\;.
\label{eq:composite}
\end{equation}
This approach allows the model to incrementally incorporate additional conditioning tasks without disrupting the knowledge acquired from previous tasks.

\section{Experimental Setup}
\label{sec:exp}

\subsection{Datasets}
\label{sec:dataset}
For the LID and ASR tasks, we utilized the ML-SUPERB benchmark~\citep{shi2023ml}. This corpus comprises two distinct data configurations: 10-minute/language and 1-hour/language for each of the 123 well-represented languages (\emph{Normal}). Additionally, both configurations include five training utterances for each of 20 selected low-resource languages (\emph{Few-shot})\footnote{We discovered that a portion of the few-shot Lithuanian (lit) training and testing data was erroneously substituted with Italian (it), leading us to omit the Lithuanian outcomes from the evaluation.}.  For the \emph{Few-shot} languages, we also considered an \emph{Extended Few-shot} condition in which we augmented the amount of data to match that of the \emph{Normal} languages. However, the Extended Few-shots only included language labels but not ASR transcripts. This aims to analyze the behavior of few-shot languages with improved language ID accuracy, as achieving satisfactory language accuracy with just five utterances is challenging. This is also a reasonable assumption since obtaining data with language labels is easier and cheaper than obtaining transcribed data.
The moderate size of this dataset was ideal for testing our approach, as it allowed us to conduct multiple ablation experiments with limited computing resources.
As ML-SUPERB lacks speaker labels, we combined it with VoxCeleb2~\citep{nagrani2017voxceleb} for training models incorporating the SV task. VoxCeleb2 contains 1,092 speech hours from 5,994 speakers, although it lacks LID labels and ASR transcripts. The SV task was tested on the VoxCeleb1 original set.
The speech was augmented with Musan noise \cite{snyder2015musan} and reverberation \cite{ko2017study} during SV training.

\subsection{Model Architecture}
\label{sec:model archi}

\paragraph{SSLR Models.}
In our system, we employed the best multilingual SSL back-bones in the ML-SUPERB benchmark: Wav2Vec2-XLSR with 300M parameters, trained on 128 languages\footnote{https://huggingface.co/facebook/wav2vec2-xls-r-300m}, and the 100M parameter multilingual Hubert (mHuBERT) model \citep{lee2021textless}, trained on English, Spanish, French data from VoxPopuli \citep{wang2021voxpopuli} unlabeled speech as our foundational acoustic encoders. These models have demonstrated their efficacy in processing a wide range of linguistic inputs and form the backbone of our system. We experimented with the S3PRL \citep{yang2021superb} and ESPnet \citep{watanabe2018espnet} toolkits. Our training dataset combined data labeled for ASR+LID labels, LID only, or SV only. Hence, we computed the losses only for the available tasks for each sample. Detailed information on the remaining training hyperparameters is provided in the appendix, and the code will be made available for reproducibility. A model's training takes about one day using 2 A100 GPUs.

\vspace{-2mm}
\paragraph{Speaker and Language Decoders.}
\label{sec:svlid_dec}
The speaker and language decoders are based on the ECAPA-TDNN architecture~\citep{desplanques2020ecapa}. Initially, a convolutional layer projects SSL representation to the decoder dimension (512 for LID and 1024 for SV). This is followed by a sequence of 1-dimensional SE-Res2Net~\citep{Gao_2021_Res2Net} layers (one for LID and three for SV).
Next, channel-wise attentive statistic pooling aggregates the frame-level features into a single utterance-level vector, which is projected into lower-dimensional speaker embedding. The training loss was Additive Angular margin-softmax~\citep{deng2018arcface} with margin=0.3 for SV and margin=0.0 for LID. Large margin helps to create highly compact speaker representations~\citep{Villalba2022}, while being detrimental in LID~\citep{villalba23_interspeech}.
The SV and LID decoders producing the final result consume a weighted average of all SSL layers. Meanwhile, the ones estimating the conditioning embeddings use a weighted average of the SSL layers evaluated up to that point in the chain. Note that all SV and LID decoders share parameters, so the number of trainable parameters remains independent of the frequency with which we re-compute the conditioning embeddings.

\vspace{-2mm}
\paragraph{ASR decoder.}
\label{sec:asr_dec}
The ASR decoder conforms to the framework set by the ML-SUPERB benchmark \citep{shi2023ml}, facilitating comparable evaluations. A convolutional downsampling layer halves the SSL feature sequence duration. These features are channeled into a two-layer Transformer with 256-dim self-attention, eight attention heads, and 1024-dim feed-forward layers. A linear output layer with connectionist temporal classification (CTC) loss predicts multi-lingual character-level tokens.

\section{Experiments and Results}
\label{sec:experiments}

\begin{table}[t]
\centering
\caption{Evaluation of adapted XLSR models on the 10-min ML-SUPERB and VoxCeleb dataset for LID, ASR, and SV tasks. These evaluations test the encoder's generalizability across different tasks, demonstrating effectiveness without further task-specific tuning.}
\label{tab:general}
\begin{subtable}{\textwidth}
\centering
\captionsetup{font=normalsize}
\caption{LID-adapted XLSR models evaluated on LID and ASR tasks.}

\label{tab:lid_asr_adapted}
\begin{tabular}{@{}lcccccc@{}}
\toprule
\multirow{2}{*}{\shortstack{\textbf{LID} \\ \textbf{Adapted.}}} & \multirow{2}{*}{\shortstack{\textbf{Bottleneck} \\ \textbf{Dims.}}}  & \multicolumn{2}{c}{\textbf{LID Acc} $\uparrow$} & \multicolumn{2}{c}{\textbf{ASR CER} $\downarrow$} \\
\cmidrule(lr){3-4} \cmidrule(lr){5-6}
& &  \textbf{Normal} & \textbf{Few-shots} & \textbf{Normal} & \textbf{Few-shots} \\ 
\midrule
XLSR & - & 89.1 & 83.9 & 29.0 & 39.0 \\
+ LID-FT & -  & 90.1 & 84.7 & 27.0 & 37.0 \\ 
+ LID-Houlsby & 256  & 90.1 & 85.3 & 23.6 & 35.1 \\
+ LID-CA-XLSR$^{L}_\text{dual}$ (ours) & 256 & \textbf{90.2} & \textbf{85.8} & \textbf{21.7} & \textbf{33.4} \\ 
\bottomrule
\end{tabular}
\end{subtable}

\begin{subtable}{\textwidth}
\centering
\captionsetup{font=normalsize}
\caption{ASR-adapted XLSR models evaluated on ASR and SV tasks.}
\label{tab:asr_sv_adapted}
\begin{tabular}{@{}lcccccc@{}}
\toprule
\multirow{2}{*}{\shortstack{\textbf{ASR} \\ \textbf{Adapted.}}} & \multirow{2}{*}{\shortstack{\textbf{Bottleneck} \\ \textbf{Dims.}}} & \multicolumn{2}{c}{\textbf{ASR CER} $\downarrow$} & \multicolumn{2}{c}{\textbf{SV}} \\
\cmidrule(lr){3-4} \cmidrule(lr){5-6}
& &  \textbf{Normal} & \textbf{Few-shots} & \textbf{EER} $\downarrow$ & \textbf{DCF} $\downarrow$ \\ 
\midrule
XLSR & -  & 29.0 & 39.0 & \underline{1.29} & \underline{0.093} \\
+ ASR-FT & - & \textbf{17.1} & \underline{32.2} & \underline{1.29} & {0.095} \\
+ ASR-Houlsby & 256 & {20.3} & {34.6} & {1.37} & {0.097} \\
+ ASR-CA-XLSR$^{L}$ (ours)  & 256 & \underline{18.6} & \textbf{31.6} & {\textbf{1.15}} & {\textbf{0.088}} \\
\bottomrule
\end{tabular}
\end{subtable}
\vspace{-0.25cm}

\end{table}

\subsection{Generalization Ability on Unseen Tasks}
\label{sec:gen}
\paragraph{Experiment Setting.} We conducted experiments to evaluate the generalization capabilities of the adapted SSLR models on LID, ASR, and SV tasks. The SSLR models were adapted for one task (either LID or ASR) and then evaluated on both the adapted task and an unseen task. 
For LID adaptation, the SSLR was trained exclusively with LID labels. We compared three setups: full fine-tuning (LID-FT), Houlsby adaptors \cite{houlsby2019parameter} (LID-Houlsby), and our proposed condition-aware approach (LID-CA-XLSR$^{L}_\text{dual}$). In this setup, we employed an additional LID decoder using the pre-trained SSLR to pre-generate language embeddings, which were then used to condition the SSLR model for a second inference pass.
For ASR adaptation, the models were trained with ASR loss using three setups: full fine-tuning (ASR-FT), Houlsby adaptors (ASR-Houlsby), and our proposed hierarchical conditioning method with TCAC layers integrated into the SSLR model with single inference (ASR-CA-XLSR$^{L}$). During ASR adaptation, the LID decoder is integrated into the SSLR model to provide conditioning features, but SV information was not included during training.

%

\paragraph{Results.} In LID adaptation (Table~\ref{tab:lid_asr_adapted}), both LID-FT and LID-Houlsby improved LID performance compared to the pre-trained SSL baseline. However, on the unseen ASR task, the fully fine-tuned SSLR encoder improved ASR CER by only 2\%, while LID-Houlsby showed limited generalization, with CER improvements of 5.4\% and 3.9\% for normal and few-shot languages, respectively. Our LID-CA-XLSR$^{L}_\text{dual}$ method achieved significantly better generalization, improving ASR CER by 7.3\% and 6.6\% for normal and few-shot languages.
In ASR adaptation (Table~\ref{tab:asr_sv_adapted}), all models enhanced ASR performance, but ASR-Houlsby and full fine-tuning degraded SV performance relative to the baseline, highlighting their limited generalization. Our ASR-CA-XLSR$^{L}$ approach not only preserved but improved SV performance, reducing EER by relative 10.9\% and DCF by 5.4\%, showcasing strong generalization to the unseen SV task.
These results demonstrate that CA-SSLR significantly outperforms full fine-tuning and standard adaptation methods in terms of generalization. By effectively leveraging conditioning information, CA-SSLR adapts across tasks while maintaining performance on unseen ones. Our proposed conditioner offers both robust adaptations on training tasks and superior generalization, making CA-SSLR a versatile and effective solution for multilingual and multispeaker speech processing.

\subsection{Condition-Aware SSLR Model}
\label{sec:CA-SSLR 1}
\paragraph{Experiment Setting.} Table~\ref{tab:Hier-SHALLi} investigates the CA-SSLR approach with hierarchical language conditioning.  The first block of the table refers to the baseline where the foundational models are frozen, while the second block ($\text{CA-XLSR}^{L}_\text{dual}$) utilizes a separate task-specific LID model to pre-generate the language embedding. The third block presents our proposed approach, where we re-estimate the language embedding every fourth or third layer ($\text{CA-XLSR}^{L}$ (4L, 3L)) within the XLSR model, not required a separate LID system. 
The experiments utilized two types of conditioners: TCAC, which incorporates attention, and a variant without attention—referred to as Channel-wise Conditioners (CC)—where the same scale and bias are applied uniformly across all time frames.
The real-time factors (RTF) as proc-time/signal-length are provided for assessing efficiency\footnote{The RTFs are computed on an NVIDIA T4 GPU.}.

\vspace{-2mm}
\paragraph{Results.} First, we observed that both $\text{CA-XLSR}^{L}_\text{dual}$ and $\text{CA-XLSR}^{L}$ systems with TCAC  (with attention) generally performed better than the CC (w/o attention) counterparts, reaffirming the benefits of the time-wise attention design.
In the second block, $\text{CA-XLSR}^{L}_\text{dual}$ slightly outperformed $\text{CA-XLSR}^{L}$ in terms of CER for both the 10-minute and 1-hour datasets. However, its real-time factor (RTF) is akin to the combined RTFs of separate LID and ASR models since it runs Wav2Vec2 twice—once for language embedding extraction and again for ASR conditioning—posing challenges for streaming applications.
On the other hand, $\text{CA-XLSR}^{L}$(CC, 3L) excelled among the three approaches, achieving a 35.9\% and 19.0\% relative improvement in Normal and few-shot languages, respectively, compared to the baseline in the 10-minute setup, and 33.5\% and 19.8\% in the 1-hour setup. LID accuracy remained comparable among the various CA-XLSR models, with a notable performance improvement from 90.9\% to 93.4\% in 1-hour setup.



\begin{table*}[t]
\centering
\caption{ASR CER (\%) and LID Acc (\%) in ML-SUPERB 10min. and 1h. sets, comparing different layers to generate the language embedding to condition the following layers. We adapt the XLSR model for LID and ASR tasks.}
\vspace{-3mm}
\label{tab:Hier-SHALLi}
\resizebox{\textwidth}{!}{
\begin{tabular}{@{}lcccccccc@{}}
\toprule
\multirow{3}{*}{\textbf{SSL Model}} & \multirow{3}{*}{\textbf{RTF}~\(\downarrow\)} & \multirow{3}{*}{\makecell{\textbf{REL.}\\ \textbf{RTF}~\(\downarrow\)}} & \multicolumn{3}{c}{\textbf{10mins}}                                & \multicolumn{3}{c}{\textbf{1hr}}                                  \\ 
\cmidrule(rl){4-6} \cmidrule(l){7-9}
 &  &  & \multicolumn{1}{c}{LID (ACC ~\(\uparrow\))} & \multicolumn{2}{c}{ASR (CER ~\(\downarrow\))} & \multicolumn{1}{c}{LID (ACC ~\(\uparrow\))} & \multicolumn{2}{c}{ASR (CER ~\(\downarrow\))} \\ 
\cmidrule(rl){4-4} \cmidrule(rl){5-6} \cmidrule(rl){7-7} \cmidrule(l){8-9}
 &  &  & Normal           & Normal           & Few-shots          & Normal            & Normal           & Few-shots           \\ \midrule
XLSR~\citep{shi2023ml}       &    0.021       & 1.00 &     66.9      &      29.2      &   40.9       &    87.9      &    22.0    &      39.3       \\
MMS-1b~\citep{shi2023findings}       &    -    & - &     84.8      &      21.3      &   30.2       &    86.1      &    18.1    &      30.8       \\

XLSR (Ours)     &    0.021      &  1.00 & \underline{89.0}           & 29.0           & 39.0             & \underline{90.9}           & 22.7           & 36.9              \\ \midrule

CA-XLSR\textsuperscript{L}\(_\text{dual}\) \text{(CC)}  & 0.037  & 1.75  & {89.0} & \underline{18.6}  & 32.2 & 90.9 &  \underline{14.1} & 31.5 \\
CA-XLSR\textsuperscript{L}\(_\text{dual}\) \text{(TCAC)}  & 0.037  & 1.75  & 89.0 & \textbf{17.8} & 31.8  & {90.9} & \textbf{13.5} & \underline{31.4}  \\  \midrule
CA-XLSR\textsuperscript{L} \text{(CC, 4L)}     &    \textbf{0.024}  & \textbf{1.17}   & \textbf{89.1 }        & 19.7           & 31.7             & 89.6          & 16.5           & 32.2              \\
CA-XLSR\textsuperscript{L} \text{(CC, 3L)}  &     \textbf{0.027}    & \textbf{1.27}      & 88.6           & 19.4           & \textbf{31.5 }            & 90.0            & 16.0           & 32.4              \\  
CA-XLSR\textsuperscript{L} \text{(TCAC, 3L)}  &     \textbf{0.027}    & \textbf{1.27}      &   {88.6}       & \underline{18.6}           & \underline{31.6}        &    \textbf{93.4}   &  15.1 & \textbf{29.6}           \\ 
\bottomrule
\end{tabular}}
\end{table*}

\begin{table*}[]
\centering
\caption{Experiments on LID and LID + SV Hierarchical Conditioning. We adapt the XLSR and mHuBERT models for LID and ASR tasks using CA-SSLR\(^L\), and for SV tasks using CA-SSLR\(^{L, S}\). Results for Normal languages with 10-min and 1-hour datasets alongside VoxCeleb SV results.}
\label{tab:hier_lid_sv}
\vspace{-2mm}
\resizebox{0.97\textwidth}{!}{
\begin{tabular}{@{}lcccccccccc@{}}
\toprule
\multirow{3}{*}{\textbf{SSL Model}} & \multirow{3}{*}{\textbf{RTF}~\(\downarrow\)} & \multirow{3}{*}{\makecell{\textbf{REL.}\\ \textbf{RTF}~\(\downarrow\)}} & \multicolumn{4}{c}{\textbf{10min. ML-SUPERB + VoxCeleb}} & \multicolumn{4}{c@{}}{\textbf{1h. ML-SUPERB + VoxCeleb}} \\ 
\cmidrule(rl){4-7} \cmidrule(l){8-11}
                    &      &         & LID & ASR &  \multicolumn{2}{c}{SV}   & LID & ASR &  \multicolumn{2}{c}{SV} \\  
\cmidrule(rl){4-5} \cmidrule(rl){6-7} \cmidrule(rl){8-9} \cmidrule(rl){10-11}
                    &      &         & ACC~\(\uparrow\) & CER~\(\downarrow\) & EER~\(\downarrow\)   & DCF~\(\downarrow\) & ACC~\(\uparrow\) & CER~\(\downarrow\)  & EER~\(\downarrow\)   & DCF~\(\downarrow\)   \\  \midrule
mHuBERT                      &    0.015  & 1.00   & 81.9 & 38.2 & 2.19 & 0.145 & 86.2 & 30.9 & 2.19 & 0.145 \\
+ FT   &  0.015  &  1.00 & 73.0 & 36.5 & 5.85 & 0.350 & \textbf{87.7} & 32.3 & 4.01 & 0.251 \\ \midrule 
CA-mHuBERT\(^L\)  \text{(CC)}          &    0.017   & 1.13   & \underline{82.0} & 31.9 & \textbf{1.77} & 0.120  & 86.1 & \underline{25.1} & \textbf{1.77} &  \textbf{0.118} \\
CA-mHuBERT\(^{L, S}\) \text{(CC)}         &    0.018  & 1.16  & \textbf{82.2} & \textbf{31.7}  & 1.79  & \textbf{0.117}  & \underline{87.3} & \textbf{24.8} &  \underline{1.78} &  \underline{0.121} \\
\midrule\midrule
XLSR                        &  0.024  &  1.00 & \underline{89.0} & 29.0 & 1.29 & 0.093 &  {90.9} & 22.7 & 1.29 & 0.093 \\
+ FT   &  0.024  &  1.00 & 81.5 & 35.6 & 7.23 & 0.353 & 83.2 & 28.7 & 6.72 & 0.330 \\ \midrule 
CA-XLSR\(^L\) \text{(CC)}            &    0.029  &    1.23 & 88.6 & 19.4 & \underline{1.11}  & \underline{0.076} & 90.0 & 16.0 & {1.02} &  0.078 \\
CA-XLSR\(^L\)  \text{(TCAC)}            &    0.029  &    1.23 & 88.6 & \underline{18.6} & 1.15 & 0.088 & \underline{93.4} & {15.1} & {1.06} & \underline{0.077} \\
CA-XLSR\(^{L, S}\)  \text{(CC)}          &   0.032  & 1.34 &\textbf{89.1}& 18.8 & \textbf{1.04}  & \textbf{0.075}  & 88.1 & \underline{15.0} &  \textbf{0.94} & \textbf{0.073} \\
CA-XLSR\(^{L, S}\)  \text{(TCAC)}            &    0.032  & 1.34  & \underline{89.0}  & \textbf{18.3} & \underline{1.11} & 0.086 & \textbf{93.5} & \textbf{14.4} & \underline{1.01} & \underline{0.077} \\
\bottomrule
\end{tabular}}
\vspace{-4mm}
\end{table*}

\subsection{Generalist Condition-Aware SSLR Model}
\label{sec:gen_}
\paragraph{Experiment Setting.}
Table \ref{tab:hier_lid_sv} presents results for general CA-SSLR models that combine Multi-lingual ASR, LID, and SV tasks. The table compares the baselines, with frozen and fine-tuned SSL models, to two different CA-SSLR Hierarchical models ($\text{CA-SSLR}^{L}$ and $\text{CA-SSLR}^{L,S}$). We further include another well-known multilingual SSLR model, mHuBERT, for a comprehensive comparison. 
The LID conditioning systems ($\text{CA-SSLR}^{L}$) are the same as from the previous section, conditioning the SSL model only on LID embeddings, with the SV decoder added on top of SSL features without further adaptation. The LID + SV conditioning system ($\text{CA-SSLR}^{L,S}$) combines both LID and SV embeddings and is jointly trained on ASR, SV, and LID losses. The intermediate LID embeddings were recomputed every three layers as the best configuration in Table~\ref{tab:Hier-SHALLi}, and SV embeddings were recomputed every six SSL layers. 
Apart from ASR CER and LID Acc on ML-SUPERB, we present SV equal error rates (EER) and detection cost function (DCF), measured at target prior probability $p=0.05$~\citep{sadjadi22b_odyssey}, on VoxCeleb1. SV performance varied depending on whether we trained the model combining 10min ML-SUPERB + VoxCeleb2 or 1h ML-SUPERB + VoxCeleb2.


\vspace{-2mm}
\paragraph{Fine-tuning Baseline.}
In the fully fine-tuning experiment, we initialized the model with pretrained ASR, LID, and SV decoders and fine-tuned for a few epochs. However, this approach resulted in suboptimal performance compared to the frozen SSLR baseline. The "FT" experiments showed degraded performance, with LID accuracy decreasing by 5.7\%, ASR CER increasing by 3.1\%, and SV EER worsening by 4.2 in absolute values on average across the four settings. This decline is unexpected, as fine-tuning typically improves performance. This suggests that simultaneous adaptation of the SSL layers to multiple tasks causes conflicting adjustments, reducing the model's robustness. Consequently, catastrophic forgetting led to worse performance compared to the baseline. Conversely, the condition-aware SSLR models exhibited superior performance comparing with the frozen baseline, indicating that training the inserted condition layers does not alter the model's behavior for downstream tasks but improves its ability to represent the input speech data.


\vspace{-2mm}
\paragraph{$\text{Language Conditioned SSLR}$.}
$\text{CA-SSLR}^{L} (\text{CC})$ notably enhanced SV performance w.r.t. the baseline, despite its encoder being solely tuned for ASR and LID tasks. For XLSR, the EER improved by 14\% and 20\% relative for the 10-min. and 1-h. sets, respectively, while DCF improved by 16-18\%. Similarly, for mHuBERT, we observed comparable enhancements, with the EER improving by 17\% in both sets and the DCF improving by 17-18\%. This demonstrates that the CA-SSLR approach offers superior generalization capabilities compared to the original pre-trained SSL encoder, delivering improved performance. $\text{CA-SSLR}^{L} (\text{TCAC})$ performance in SV is comparable to its non-attention counterpart with better performance in LID and ASR as discussed in Sec.\ref{sec:CA-SSLR 1}.


\paragraph{$\text{Language and Speaker Conditioned SSLR}$.}
Adding a speaker conditioner to $\text{CA-SSLR}^{L,S}$ further improved its performance.  
In ASR tasks, incorporating the speaker conditioner to $\text{CA-XLSR}^{L,S}$ reduced CER by 3.1\% for the 10-min. set and 6.2\% for the 1-hr set, relative to $\text{CA-XLSR}^{L} $. 
For LID task, $\text{CA-SSLR}^{L,S} $ shows similar performance to other models with relative differences below 3\%.
For SV, $\text{CA-XLSR}^{L,S}$ using channel-wise conditioner (CC) reduced EER by 19.4-27.1\%, outperforming $\text{CA-XLSR}^{L}$. Switching from CC to TCAC yielded additional gains in ASR, adding a relative improvement of 2.7-4.0\%. In contrast, its impact on SV was more modest, with improvements in EER by 14.0-21.7\%. Overall, TCAC demonstrated better adaptation ability, while CC excelled in generalization.


\paragraph{$\text{ASR and RTF Discussion}$.}
Generally, we observed the largest improvement for ASR when including the language conditioner,  as it enables the system to adapt to produce output tokens in the correct language. Conversely, adapting the model to the input speaker provided fewer ASR gains. The XLSR model benefitted from our approach better than mHuBERT, possibly because mHuBERT is $3\times$ smaller than XLSR, but more importantly, because mHuBERT was trained on just four languages compared to 128 in XLSR. Therefore, the pre-trained mHuBERT has not encountered enough diversity in terms of languages and speakers, thereby limiting its performance in multi-lingual ASR and SV.
In terms of RTF, while the conditioned models are 13-34\% slower compared to sharing the pre-trained SSL encoder for the three tasks, both $\text{CA-SSLR}^{L}$ and $\text{CA-SSLR}^{L,S}$ offer superior performance while being much faster than running task-specific models separately, indicating a more efficient use of computational resources while running the generalist model.

\begin{figure}[t]
\begin{minipage}{0.42\textwidth}
    \centering
    \captionof{table}{Ablation study of condition-aware settings for ASR-adapted XLSR models on 10-min ML-SUPERB dataset, using CC or TCAC. Conditioning is based on predicted language labels or LID embeddings, except in the ground truth (G.T.) experiment.}
    \label{tab:asr_adapted}
    \resizebox{0.99\textwidth}{!}{ 
    \begin{tabular}{@{}lcc@{}}
        \toprule
        \multirow{2}{*}{\shortstack{\textbf{ASR} \\ \textbf{Adapted.}}} & \textbf{Normal} & \textbf{Few-shots} \\
         & \textbf{CER}~\(\downarrow\) & \textbf{CER}~\(\downarrow\) \\ \midrule
        XLSR & 29.0 & 39.0 \\ 
        + G.T. CC    & 17.2 & 27.9 \\ 
        \midrule
        + Hard CC  & 19.8 & \textbf{28.6} \\
        + Soft CC & 19.3 & 32.5 \\
        + Embed CC    & \underline{18.6} & 32.2 \\
        + Embed TCAC  & \textbf{17.8} & \underline{31.8} \\
        \bottomrule
    \end{tabular}}
\end{minipage}
\hfill
\begin{minipage}{0.55\textwidth}
    \centering
    \includegraphics[width=\textwidth]{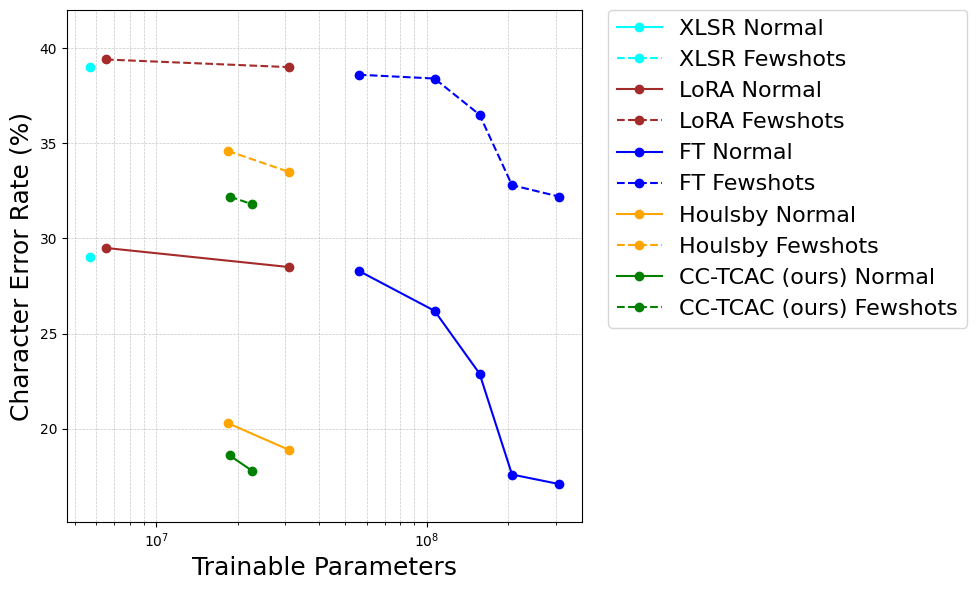}
    \caption{CER versus trainable parameters on XLSR model for Normal and Few-shots languages, demonstrating the adaptation ability for the TCA conditioner.}
    \label{fig:params}
\end{minipage}
\vspace{-5mm}
\end{figure}

\subsection{Analysis of the TCA Conditioner}
\paragraph{Ablation study of Conditioning Approach.}
\label{sec:Abl}
Table \ref{tab:asr_adapted} conducts an ablation study for different conditioning methods with CA-XLSR$^L_\text{dual}$ settings within the ML-SUPERB 10min dataset regarding ASR CER.
First, we used conditioners without attention (CC) on the ground truth LID predictions (G.T.), serving as the upper bound for the performance of our proposed approach. This improved the \emph{Normal} languages from 29.0\% to 17.2\%, and \emph{Few-shot} languages from 39.0\% to 27.9\%, w.r.t. the pre-trained XLSR model. This showcases the potential of the condition-aware SSLR. Following, we compared conditioning on hard-predicted language labels (\emph{Hard}), soft-predicted language labels (Soft), and language embeddings from the LID decoder bottleneck layer (Embed) for comparison. Conditioning on \emph{Hard} LID labels improved the most in \emph{Few-shot} languages, improving by 26\% relative to the baseline. On the other hand, the \emph{Embed} case outperformed the \emph{Soft} case and provided balanced performance for both  \emph{Normals} and \emph{Few-shots} languages. Additionally, we compared CC to TCAC. The TCAC provided the best overall results, improving \emph{Normals} and \emph{Few-shots} by 38.6\% and 18.5\%, respectively, w.r.t. baseline. 

\paragraph{Parameter Efficiency in CER Reduction.}
Figure~\ref{fig:params} compares CER versus the number of trainable parameters for different adaptation methods, including our proposed Channel-wise Conditioner and Time-Channel Attention Conditioner (CC-TCAC), the Houlsby adapter, LoRA~\citep{hu2021lora}, full fine-tuning (FT), and the baseline XLSR model.
The Houlsby adapters, with hidden dimensions of 256 and 512, have 18.4M and 30.9M trainable parameters. In comparison, the CC-TCAC approach, conditioned on precomputed LID embeddings with 256 dimensions (18.7M for CC and 22.6M for TCAC), achieves lower CERs with similar parameter counts.
LoRA provided only marginal gains over the baseline, aligning with findings from~\cite{chen2023exploring}. In contrast, FT required fine-tuning 16–24 layers (200–300M parameters) to achieve comparable CER reductions, making CC-TCAC about ten times more efficient.
As discussed in Sec~\ref{sec:gen}, CC-TCAC's key contribution is its superior generalization ability.
While the Houlsby adapter enhances task-specific adaptation, it falls short in generalizing to unseen tasks. In contrast, CC-TCAC achieves both effective adaptation and robust generalization, making it a versatile solution for diverse applications.

\section{Conclusion}
\label{sec:conclusion}
This paper introduces the CA-SSLR framework, an innovative approach that integrates conditioning into pre-trained Self-Supervised Learning (SSL) models by adapting only the trainable conditioner. Through a hierarchical self-conditioning mechanism, where intermediate language and speaker features condition the upper layers of the SSL model, CA-SSLR achieves a 33\% reduction in Character Error Rate compared to the pre-trained baseline, matching the performance of single-task fully fine-tuned models. Additionally, it improves Speaker Verification EER by 27\% and reduce Language Identification errors by relative 10\% in average. The results indicate that condition-aware SSLR models enhance the model's interpretation of input speech data, providing superior performance compared to traditional fine-tuning methods. This improvement is achieved by dynamically tailoring the model’s response to the input language and speaker characteristics, ensuring robust generalization across various tasks. In summary, CA-SSLR offers a versatile and efficient approach to integrating conditioning information into pre-trained models. This method not only enhances performance across multiple tasks but also ensures efficient parameter utilization, supported by an improved RTF that facilitates its application in real-world scenarios.


\paragraph{Broader Impact and Limitations}
The CA-SSLR methodology improves the conditioning of pre-trained Self-Supervised Learning (SSL) models for speech processing, improving performance with minimal fine-tuning and reducing computational resource requirements. This advancement facilitates the deployment of robust models in resource-constrained environments, promoting broader access to advanced speech technology.
However, there are potential risks. The conditioning mechanisms might amplify biases in the training data, leading to unfair outcomes, particularly for underrepresented languages and speaker groups. Ensuring diverse and balanced datasets, along with continuous monitoring, is crucial to mitigate these risks and prevent perpetuating existing inequities.

\bibliographystyle{plainnat}
\bibliography{custom}

\newpage

\appendix

\section{Model/Dataset Details and Training Hyper-parameters}
\label{sec:model detail}
This appendix provides detailed configurations and hyper-parameters for the decoder models used in our experiments, including ASR, LID, SV decoders, and the CA-SSLR models. The rationale behind specific hyper-parameter choices and architectural details are also discussed to offer insights into the experimental setup and model optimization strategies.

\subsection{Decoder Models for ASR, LID, and SV}

The ASR, LID, and SV decoder models were optimized for their respective tasks through careful selection of hyper-parameters and architectural configurations. The ASR model directly follows the setting in ML-SUPERB benchmark \citep{shi2023ml} for comparison. Table \ref{tab:decoder_models} summarizes these configurations. The ``full'' means one epoch is trained by passing all the training data.

\begin{table}[htbp]
\centering
\caption{Hyper-parameters used for training ASR, LID, and SV decoder models}
\label{tab:decoder_models}
\begin{tabular}{@{} lccc @{}}
\toprule
 & ASR & LID & SV \\
\midrule
Feature Projection & 80 & 80 & 80 \\
Decoder Layers & 2 & 1 & 3 \\
Hidden Channels & 256 & 512 & 1024 \\
Dropout Rate & 0.1 & 0.3 & 0.0 \\ \hline
Loss & CTC & CE & CE \\
Learning Rate & 0.0001 & 0.0001 & 0.001 \\
Warmup Steps & - & - & 1000 \\
Effective Batch Size & 32 & 128 & 512 \\
Iterations per Epoch & 5000 & 5000 & full \\
Epochs & 20 & 10 & 20 \\
\bottomrule
\end{tabular}
\end{table}

\begin{table*}[htbp]
\centering
\caption{Training hyper-parameters for CA-SSLR models. The superscripts ``Dec'' and ``Feat'' represent the decoder and the feature projection layer, respectively.}
\label{tab:shalli_models}
\begin{tabular}{@{}lccc@{}}
\toprule
 & \multicolumn{2}{c}{\(\text{CA-SSLR}^{L}\)} & \(\text{CA-SSLR}^{L,S}\) \\
\midrule
Training Data & ML-SUPERB & VoxCeleb & ML-SUPERB + VoxCeleb \\
Condition Embedding & \multicolumn{2}{c}{256 (L)} & 256 (L) + 256 (S) \\
Condition Dropout Rate & \multicolumn{2}{c}{0.5}  & 0.5 \\ \midrule
Initialization & \multicolumn{2}{c}{\makecell[c]{ASR\textsuperscript{Dec}, LID\textsuperscript{Dec}, \\ and SV\textsuperscript{Dec}}} & \(\text{CA-SSLR}^{L}\) \\ \midrule
Trainable Modules & \makecell[c]{ASR\textsuperscript{Dec}, LID\textsuperscript{Feat}, \\ and Adapters}  & SV\textsuperscript{Dec}  & \makecell[c]{ASR\textsuperscript{Dec},  LID\textsuperscript{Feat}, \\ SV\textsuperscript{Feat}, and Adapters} \\ \midrule
Loss & ASR + LID & SV & ASR + LID + SV \\
Learning Rate & 0.0001 & 0.001  & 0.0001 \\
Effective Batch Size & 32 & 512  & 32 \\
Iterations per Epoch & 5000 & full &  5000 \\
Epochs & 20 & 20 & 20 \\
\bottomrule
\end{tabular}
\end{table*}

\subsection{CA-SSLR Hierarchical Models}
\label{sec:append_mhubert}

Table \ref{tab:shalli_models} provides detailed configurations for the CA-SSLR model. In the $\text{CA-SSLR}^{L,S}$ setup, two 256-dimensional embeddings are used to encapsulate language (L) and speaker (S) information, which then determine the parameters ($\alpha_{\text{L}}$, $\gamma_{\text{L}}$, $\beta_{\text{L}}$) and ($\alpha_{\text{S}}$, $\gamma_{\text{S}}$, $\beta_{\text{S}}$) following the procedure outlined in Eq.~\ref{eq:composite}. 
The training adopts a stepwise approach, using initial parameters from an earlier phase to set up the next. The pretrained ASR, LID, and SV decoders serve as the foundation for initializing $\text{CA-SSLR}^{L}$; the SV decoder is fine-tuned further on top of $\text{CA-SSLR}^{L}$; and both $\text{CA-SSLR}^{L}$ and fine-tuned SV decoder initialize $\text{CA-SSLR}^{L,S}$.  In the table's ``Trainable modules'' row, the notations LID\textsuperscript{Feat} and SV\textsuperscript{Feat} indicate that the feature projection layers of the LID and SV decoders are adjustable during the training process. We conduct the in Table \ref{tab:shalli_models} and Figure \ref{fig:params} multiple times, and the variation are all within 0.2\% CERs range.

\subsection{Dataset License and Details}
\label{sec:license}
\subsubsection{ML-SUPERB Dataset}
The ML-SUPERB dataset is assembled from a wide collection of multilingual speech corpora, with each contributing corpus being governed by one of a variety of open-source licenses, such as Creative Commons, MIT, GNU, or Free-BSD. These licensing agreements guarantee that the dataset is openly available and can be used freely for both commercial and scholarly research purposes. 
The 10-minute training set encompasses 37.4 hours of data, and the 1-hour dataset increases the total to 222.4 hours of data. Additionally, the dataset includes development and testing sets, containing 41.8 hours and 45.0 hours of data, respectively. This dataset is designed for multilingual speech recognition and language identification, as used in our work.

In the original ML-SUPERB settings, there are two types of languages:

\begin{itemize}
    \item {Normal Languages}: Each has 10 minutes or 1 hour of data per language, used for both LID and ASR training with transcriptions.
    \item {Few-Shot Languages}: Each has only 5 utterances. In the original settings, these languages are not presented in the results for LID training and are used for ASR training with available transcriptions.
\end{itemize}

For the {extended few-shot condition}, we incorporate the language labels from these few-shot data for LID training but continue using only 5 utterances with transcriptions for ASR training. Since language labels are more accessible than transcriptions, especially in low-resource scenarios. Table~\ref{tab:extended_few_shot} summarizes the data configurations for the original and extended few-shot conditions.

\begin{table}[h]
    \centering
    \caption{Data configurations for the original and extended few-shot conditions in ML-SUPERB.}
    \label{tab:extended_few_shot}
    \begin{tabular}{lccc}
        \toprule
        \textbf{Data Per Language} & \textbf{Language Type} & \textbf{LID Training} & \textbf{ASR Training} \\
        \midrule
        \multirow{2}{*}{Original Settings} & Normal & 10 min -- 1 hr & 10 min -- 1 hr \\
        & Few-Shot & Not presented in result & 5 utts \\ \midrule
        Extended Few-Shot & Few-Shot & 10 min -- 1 hr (language labels only) & 5 utts  \\
        \bottomrule
    \end{tabular}
\end{table}

\subsubsection{VoxCeleb Dataset}
The VoxCeleb dataset is available under the Creative Commons Attribution 4.0 International license and encompasses comprehensive training, development, and testing data collection. Specifically, it contains 1092 hours of audio from 5,994 speakers for training, 110 hours from 4,933 speakers for development, and 20 hours from 40 speakers designated for testing. Designed to facilitate speaker verification and identification tasks, aligns with our usage in the speaker verification task. To ensure privacy, speaker names within the dataset are anonymized and represented through unique speaker IDs.

\section{CER vs. Trainable Parameters}
Table \ref{tab:mhubert} compares the mHuBERT model's ASR performance against the number of trainable parameters, where the XLSR counterpart is shown in Fig.~\ref{fig:params}. Both $\text{CA-mHubert}^{L}_\text{dual}$ and $\text{CA-mHubert}^{L,S}_\text{dual}$ are with 256 condition feature dimensions. Notably, the $\text{CA-mHubert}^{L}_\text{dual}$ model excels in few-shots scenarios, while the $\text{CA-mHubert}^{L,S}_\text{dual}$ yields CERs for normal languages comparable to a fully fine-tuned 12-layer mHuBERT model using only 15.9M parameters. This efficiency demonstrates the TCA conditioner's capability in the CA-SSLR framework to deliver fine-tuned levels of ASR accuracy with a significantly reduced parameter count, providing an optimal balance for practical ASR applications.

\begin{table}[h]
    \centering
\caption{Comparison of trainable parameters and CERs on ML-SUPERB 10min dataset, including fine-tuning top layers, LoRA, and dual-inference condition-aware mHuBERT model.}
\label{tab:mhubert}
\begin{tabular}{@{}lccc@{}}
\toprule
\multirow{2}{*}{\textbf{Approach}} & \multirow{2}{*}{\makecell{\textbf{Trainable}\\ \textbf{Params.}}} & \textbf{Normal} & \textbf{Few-shots} \\
 & & \textbf{CER}~\(\downarrow\) & \textbf{CER}~\(\downarrow\) \\ \midrule
mHuBERT       & 5.7M    & 38.2 & 42.9  \\ \midrule
LoRA          & 15.2M   & 38.3 & 42.7  \\ \midrule
FT (2L)       & 19.9M   & 36.0 & 42.3  \\
FT (4L)       & 34.0M   & 35.0 & 42.3  \\
FT (6L)       & 48.2M   & 33.1 & 41.4  \\
FT (8L)       & 62.6M   & 32.0 & 40.5  \\
FT (12L)      & 90.8M   & \textbf{30.8} & 40.5 \\ \midrule
\(\text{CA-mHuBERT}^{L}_\text{dual}\)    & 10.8M   & 31.6 & \textbf{40.0} \\
\(\text{CA-mHuBERT}^{L,S}_\text{dual}\)  & 15.9M   & \textbf{30.9} & 40.4 \\ \bottomrule
\end{tabular}
\end{table}

\section{Training Efficiency and Resource Usage}

We compare the training speed and resource consumption of different adaptation methods, including Houlsby Adapters, CA-SSLR, and full fine-tuning (FT). Table~\ref{tab:training_efficiency} summarizes the bottleneck dimensions, training times, and peak memory usage for each method. We evaluate the training speech for 10k iterations with batch size 8. We found that the CA-SSLR approach ranks second compared to the Houlsby Adapter and a fully fine-tuning approach in speed and memory usage. 
However, it is important to note that CA-SSLR surpasses the Houlsby Adapter in adaptation effectiveness and generalization ability, as demonstrated in Table~\ref{tab:general}. These results indicate that although CA-SSLR incurs a moderate increase in training resources, it provides benefits in performance and generalization. We acknowledge that the current implementation of CA-SSLR is not yet optimized for speed and memory efficiency. Future work will focus on optimizing the model to reduce training time and memory consumption without compromising performance.

\begin{table}[h]
    \centering
    \caption{Comparison of adaptation methods in terms of bottleneck dimensions, training speed, and peak memory usage.}
    \label{tab:training_efficiency}
    \begin{tabular}{@{}lccc@{}}
        \toprule
        \textbf{Method} & \textbf{Bottleneck Dims.} & \textbf{Training Speed} & \textbf{Peak Memory Usage} \\ \midrule
        Houlsby Adapter & 256 & 76 mins & 58 GB \\
        CA-SSLR$^L$ (3L) & 256 & 120 mins & 68 GB \\
        Fine-Tuning (FT) & - & 135 mins & 79 GB \\ \bottomrule
    \end{tabular}
\end{table}

\section{RTF Analysis}
Tables~\ref{tab:rtf} and~\ref{tab:rtf_total} present the real-time factor (RTF) for each individual component, as well as for the combined systems. In Table~\ref{tab:rtf_total}, the term "separated tasks" refers to duplicating and fine-tuning the SSLR for each task individually, along with the corresponding total RTF.

\begin{table}[h!]
\centering
\begin{tabular}{l|c}
\hline
\textbf{Block} & \textbf{RTF} \\
\hline
ASR Decoder & 0.004 \\
LID Decoder & 0.001 \\
SPK Decoder & 0.003 \\ \hline
XLSR SSL & 0.016 \\ 
CA-XLSR$^{S}$ (6L) & + 0.003 \\
CA-XLSR$^{L}$ (4L) & + 0.004 \\
CA-XLSR$^{L}$ (3L) & + 0.006 \\ \hline
mHubert SSL & 0.007 \\ 
CA-mHubert$^{S}$ (6L) & + 0.001 \\
CA-mHubert$^{L}$ (3L) & + 0.002 \\
\hline
\end{tabular}
\caption{RTF for different components.}
\label{tab:rtf}
\end{table}

\begin{table*}[h!]
\centering
\begin{tabular}{lclc}
\toprule
XLSR Approach & RTF & mHubert Approach & RTF \\
\midrule
\multicolumn{3}{c}{ASR + LID (Table 2)} \\
\midrule
XLSR + ASR + LID & 0.021 & mHubert + ASR + LID & 0.013 \\
CA-XLSR$^{L}$(4L) & 0.024 & - \\
CA-XLSR$^{L}$(3L) & 0.027 & CA-mHubert$^{L}$(3L) & 0.015 \\
Separated 2 tasks (+XLSR) & 0.037 & Separated 2 tasks (+mHubert) & 0.020 \\
\midrule
\multicolumn{3}{c}{ASR + LID + SV (Table 3)} \\
\midrule
XLSR + ASR + LID + SV & 0.024 & mHubert + ASR + LID + SV & 0.015 \\
CA-XLSR$^{L}$ & 0.029 & CA-mHubert$^{L}$ & 0.017 \\
CA-XLSR$^{L,S}$ & 0.032 & CA-mHubert$^{L,S}$ & 0.018 \\
Separated 3 tasks (+2*XLSR) & 0.055 & Separated 3 tasks (+ 2*mHubert) & 0.030 \\
\bottomrule
\end{tabular}
\caption{Total RTFs for combined systems.}
\label{tab:rtf_total}
\end{table*}

\begin{table*}[ht!]
\centering
\caption{Evaluation of LID and ASR performance in terms of Accuracy (Acc) and Character Error Rates (CERs) for few-shot learning in low-resource languages using the ML-SUPERB 10-minute set, comparing XLSR and mHuBERT models.}
\label{tab:language_accuracy}
\begin{tabular}{@{}lcccccccc@{}}
\toprule
\textbf{Lang.} & \multicolumn{2}{c}{XLSR} & \multicolumn{2}{c}{\(\text{CA-XLSR}^{L,S}\)} & \multicolumn{2}{c}{mHuBERT} & \multicolumn{2}{c}{\(\text{CA-mHuBERT}^{L,S}\)} \\ 
\cmidrule(r){2-3} \cmidrule(r){4-5} \cmidrule(r){6-7} \cmidrule(r){8-9}
 & \textbf{Acc}~\(\uparrow\) & \textbf{CER}~\(\downarrow\) & \textbf{Acc}~\(\uparrow\) & \textbf{CER}~\(\downarrow\) & \textbf{Acc}~\(\uparrow\) & \textbf{CER}~\(\downarrow\) & \textbf{Acc}~\(\uparrow\) & \textbf{CER}~\(\downarrow\) \\ 
\midrule
bos  & \textbf{82.0} & 21.3 & 70.0 & \textbf{11.7} & \textbf{30.0} & 29.0 & 28.0 & \textbf{26.0} \\
ceb  & 97.6 & 20.5 & \textbf{97.6} & \textbf{12.4} & 92.9 & 27.2 & \textbf{97.6} & \textbf{25.4} \\
dan  & \textbf{89.5} & 44.7 & 76.5 & \textbf{37.0} & 80.1 & 49.1 & \textbf{80.4} & \textbf{47.5} \\
epo  & \textbf{81.7} & 15.3 & 76.9 & \textbf{14.5} & 46.2 & \textbf{24.8} & \textbf{52.9} & 25.4 \\
frr  & 87.5 & 33.6 & \textbf{89.3} & \textbf{29.9} & \textbf{67.9} & 40.4 & 63.4 & \textbf{37.7} \\
ful  & 55.0 & \textbf{27.1} & \textbf{67.5} & 28.2 & \textbf{37.5} & 34.7 & 62.5 & \textbf{32.0} \\
kaz  & 98.0 & 32.6 & \textbf{99.3} & \textbf{21.5} & \textbf{91.4} & 37.8 & 88.7 & \textbf{36.0} \\
kea  & 84.1 & 28.9 & \textbf{90.9} & \textbf{27.6} & 65.9 & 35.2 & \textbf{75.0} & \textbf{33.1} \\
lit  & 87.3 & 52.0 & \textbf{87.7} & \textbf{45.2} & 79.3 & 52.4 & \textbf{82.5} & \textbf{49.3} \\
luo  & \textbf{100.0} & 29.4 & 95.1 & \textbf{24.4} & 92.7 & \textbf{29.4} & \textbf{92.7} & 30.1 \\
srp  & \textbf{64.8} & 57.4 & 50.9 & \textbf{48.1} & \textbf{53.5} & 56.7 & 45.7 & \textbf{56.2} \\
sun  & 93.5 & 26.4 & \textbf{94.4} & \textbf{19.1} & \textbf{94.4} & 32.7 & 93.5 & \textbf{26.2} \\
tok  & 98.5 & 15.4 & \textbf{98.5} & \textbf{13.1} & 98.5 & 23.2 & \textbf{98.5} & \textbf{19.2} \\
tos  & \textbf{100.0} & 49.1 & 99.4 & \textbf{44.7} & 99.4 & 53.1 & \textbf{99.4} & \textbf{48.9} \\
tso  & \textbf{84.0} & 25.4 & 81.3 & \textbf{21.7} & \textbf{83.3} & 29.4 & 81.3 & \textbf{26.0} \\
tsn  & \textbf{87.1} & 22.7 & 85.7 & \textbf{17.3} & 83.6 & 27.3 & \textbf{84.3} & \textbf{23.9} \\
tur  & \textbf{82.4} & 60.0 & 79.1 & \textbf{37.0} & \textbf{62.6} & 65.1 & 57.7 & \textbf{61.7} \\
umb  & \textbf{64.0} & 24.6 & 40.0 & \textbf{23.3} & \textbf{44.0} & \textbf{29.8} & 36.1 & 30.1 \\
vie  & \textbf{94.7} & 88.4 & 92.9 & \textbf{83.1} & \textbf{85.8} & \textbf{80.1} & 74.2 & 80.3 \\
zul  & 60.6 & 20.4 & \textbf{62.3} & \textbf{14.4} & \textbf{53.7} & 24.3 & 52.0 & \textbf{20.4} \\
\bottomrule
\end{tabular}
\end{table*}

\section{Few-shots Results}
Within the ML-SUPERB dataset's 20 few-shots languages, we examined the performance of CA-SSLR against the established SSL baselines, XLSR and mHuBERT, on models trained in the 10-minute ML-SUPERB set. The LID results indicate a close match between CA-SSLR and the baseline, with approximately half of the few-shot languages exhibiting improvements or matching their baseline performance. Section~\ref{sec:Abl} reveals that SSL-based LID models are inherently effective, and extending full fine-tuning does not necessarily enhance results. This observation aligns with the outcomes of other classification tasks adeptly handled by SSL models, as documented in~\citep{chen2023exploring}. 
Furthermore, the CA-SSLR framework demonstrates subtle enhancements for the Normal languages in the 10-minute set in Table~\ref{tab:hier_lid_sv}, indicating that the LID performance remains robust despite the encoder's additional modifications. 

Regarding the ASR results, most languages achieved significant CER reductions, ranging from a modest few percent to over 30\%, when compared with SSL baselines. Notably, the Bosnian (bos) language experienced an impressive 45.1\% relative improvement in CER, while Cebuano (ceb) improved by 39.5\% with the XLSR model. With the mHuBERT model, the most substantial gains were observed in Sundanese (sun) and Toki Pona (took), with 19.9\% and 17.2\% CER relative improvements, respectively. These results underscore the CA-SSLR framework's profound effect in bolstering ASR performance for few-shot languages, especially demonstrating more pronounced improvements with the XLSR model.

When examining the correlation between LID accuracy and ASR performance, it is apparent that a lower CER does not necessarily align with high LID accuracy. For instance, Serbian (srp) on the XLSR model, despite having a modest LID accuracy of 50.9\%, shows a CER improvement from 57.4\% to 48.1\%. Conversely, Fulah (ful), the sole language to exhibit a CER increase in the XLSR model, presents a higher LID accuracy of 67.5\%. This indicates that the CA-SSLR framework's efficacy is not solely contingent on high LID prediction accuracy. CA-SSLR's reliance on embeddings instead of one-hot hard labels for predictions enables the model to maintain or improve ASR performance despite suboptimal LID scores. This approach allows the model to utilize embeddings to distinguish between easily confused languages, enabling the ASR model to predict the correct language accurately.

\section{Decode Examples}
Table~\ref{tab:asr_text} visualizes ASR outcomes for the XLSR and $\text{CA-SSLR}^{L,S}$ models on the ML-SUPERB 10-minute dataset, covering both few-shot and standard language scenarios. It highlights CA-SSLR's superior language recognition capabilities and success in rectifying the misclassifications encountered with XLSR, often resulting in completely incorrect transcriptions. This is evident in languages such as Lithuanian and Turkish, categorized as few-shot, and Bulgarian, which is better resourced (normal). These findings demonstrate the TCA conditioner's effectiveness in accurately managing LID embedding features and distinguishing between languages for downstream tasks.

Moreover, the results from other samples suggest that CA-SSLR can achieve better outcomes during training due to its incorporation of language information, even when the XLSR model correctly predicts the language. This underscores the efficacy of the TCA conditioner in exploiting language-specific data, thereby enabling CA-SSLR to achieve heightened accuracy across a diverse range of languages.

\section{Ethical Statement}
\label{sec:ethic}

We affirm our commitment to ethical research practices, including respect for privacy and the responsible use of data. The proposed CA-SSLR model improves multi-lingual ASR in 143 languages, including 20 low-resource ones with just five training utterances. In this manner, CA-SSLR contributes to the democratization of speech technology, fostering inclusivity for previously underserved communities. Furthermore, CA-SSLR prioritizes the reduction of computational costs at evaluation time, thereby aiming to mitigate the environmental impact associated with speech applications. We utilized publicly available datasets, namely ML-SUPERB and VoxCeleb, chosen for their moderate size to minimize computing requirements. Our utilization of pre-trained models, specifically XLSR and mHuBERT, aligns with their intended research purposes, as they are widely used within the speech research community.

However, the capacity for conducting speech and speaker recognition in human conversations poses a notable ethical concern linked to covert eavesdropping by nefarious entities. This capability could be exploited by authoritarian government bodies seeking to suppress free speech and identify dissidents.  Therefore, it is imperative to promote public awareness and comprehension regarding the automated analysis of spontaneous speech and its ramifications.

\begin{table*}[h!]
  \centering
  \begin{tabularx}{\textwidth}{|c|c|X|X|X|}
    \hline
    Language & Group  & {Ground Truth} & {XLSR} & \(\text{CA-XLSR}^{L,S}\)\\
    \hline
\makecell[c]{Esperanto \\ (epo) } & Few-shots &  LI STUDVOJAĜIS AL ITALIO HISPANIO KAJ FRANCIO & 
    L\substitution{ES}ST\insertion{O}\substitution{S} VO\insertion{L}JA\substitution{G}IS A\deletion{L} \insertion{L}ITALIO HISPANIO \insertion{C}KA\substitution{I} FRANCIO & 
    LI STUD VOJA\substitution{G}IS AL ITALIO HISPANIO KAJ FRAN\insertion{Z}CIO \\
    \hline 
\makecell[c]{Lithuanian \\ (lit) }& Few-shots &  KARALIUS NEIŠDRĮSO KALDINTI VARINIŲ MONETŲ KARALIŠKOJOJE MONETŲ KALYKLOJE & 
    
    \textcyr{\substitution{КАРАЛЮС}S \substitution{НЕ} \substitution{ЖДРИСА} \insertion{КА}\substitution{ЛЬНЕН } \substitution{ТЕ ВОАРИНУ МОНЕТУ КАРАЛЮШКО}J\deletion{O}\substitution{A}E \insertion{Б} \substitution{МОНАТУ} \substitution{КАЛІКЛО}J\substitution{A}}
    &
    KARAL\substitution{J}US NE I\substitution{Ž}DR \substitution{I} SO KAL\substitution{NE}N T\deletion{I}\insertion{E} V\substitution{O}RIN\insertion{I}\substitution{U} MONET\substitution{U} KARAL\insertion{J}\substitution{U}ŠKOJO JE MONET\substitution{U} KAL\substitution{I}KLOJE  \\

    \hline 
\makecell[c]{Serbian \\ (srp) }& Few-shots &  OVO OTKRIĆE TAKOĐE PRUŽA UVID U EVOLUCIJU PERA KOD PTICA &  OVO O\substitution{D}KRI\insertion{J}\insertion{Č}\substitution{I}E TAKO\insertion{ }\substitution{Ć}E PRUŽA UVID\deletion{ }U\substitution{J}EVOLUC \deletion{I} JU PERA\deletion{ }KO\substitution{B} \deletion{P}TI\deletion{C}\insertion{T}
 & OVO O\substitution{D}\insertion{ }KRI\insertion{Č}ĆE TAKO ĐE PRUŽA UVID\deletion{ }U EVOLUCI JU PERA KOD \deletion{P}TI\deletion{C}\substitution{T}\\
    \hline 
    
\makecell[c]{Northern \\ Frisian  \\ (frr) } & Few-shots &  MEI IK TAKOM WIKE DAT BOEK FAN DY LIENE & M\deletion{E}\substitution{A} \substitution{E}K TA\insertion{KG}KO\insertion{M}M\insertion{E} WIG\substitution{G}E DAT BO\deletion{E}K \substitution{V}AN DIE \insertion{I}\substitution{E} \deletion{L}IE\insertion{Ë}NE
 & MA\substitution{A} IK TAKO\insertion{M}M\insertion{E} WIKE DAT BOEK \substitution{V}AN \insertion{I}\substitution{E} LIENE
\\
    \hline 

\makecell[c]{Turkish \\ (tur) }& Few-shots & TÜM BUNLAR ILGIMIZI ÇEKSE DE UZUN KALAMAZDIK  & \textcyr{\substitution{Т}\substitution{Ү}\substitution{М} \substitution{Б}\substitution{У}\substitution{Н}\substitution{Л}\substitution{А}\substitution{Р} \substitution{И}\substitution{Л}\substitution{Г}\substitution{И}\substitution{М}\substitution{З}\substitution{Е}\substitution{Й} \substitution{Ч}\substitution{А}\substitution{К} \substitution{С}\substitution{Е} \substitution{Д}\substitution{Е} \substitution{У}\substitution{З}\substitution{У}\substitution{Н} \substitution{К}\substitution{А}\substitution{Л}\substitution{А}\substitution{М}\substitution{А}\substitution{С}\substitution{Д}\substitution{Ы}\substitution{К}
} &  TÜM BUNLAR İLGİMİZİ Ç\substitution{\textschwa}KS\deletion{E}\substitution{\textschwa} DE UZUN KALAMAZDIK
\\ \hline
\makecell[c]{Belarusian \\ (bel) }  & Normal &\textcyr{НА ПЕРШЫМ ПЛАНЕ КАРЦІНЫ НАМАЛЯВАНЫ ГОСЦІ НАКРЫТЫЯ ПЯЛЁСТКАМІ РУЖ} & \textcyr{НА ПЕРШ\substitution{И}М ПЛАНЕ КАРЦ\substitution{И}Н\substitution{Е} НАМАЛ\substitution{Е}ВАНЫ\insertion{Й} ГОС\substitution{ТЕ} НА \insertion{ }КРЫТ\deletion{Ы}\substitution{Е} П\substitution{Е}ЛЁСТК\substitution{Ы}М\deletion{І}\substitution{Е} РУЖ
}

 &  \textcyr{НА ПЕРШЫМ ПЛАНЕ КАРЦІНЫ НАМАЛЯВАНЫ\insertion{Й} ГОСЦІ НАКРЫТЫЯ П\substitution{Е}ЛЁСТКАМІ РУЖ} \\ \hline
 
\makecell[c]{Bulgarian \\ (bul) }  & Normal  & \textcyr{СЛЕД МАЛКО ВАСИЛЕНА ПАК ИЗЛЕЗЕ} & \substitution{SLED MAUKU VASILENA} \insertion{P}\substitution{AKI} \textcyr{\deletion{И}}\substitution{ZLEZE}
 & \textcyr{СЛЕД МАЛКО ВАС\substitution{Е}ЛЕНА ПАК ИЗЛЕЗЕ} \\ \hline
 
\makecell[c]{Basque \\ (eus) }  & Normal  & KORRONTE ETIKO HORREN HELBURUA ZORIONTASUNA LORTZEA DA & KO\deletion{R}RONTE\deletion{ }ETIKO \deletion{H}O\deletion{R}REN\deletion{ }\deletion{H}ELBURUA ZORIO\insertion{Ą}NTASUNA LOR\deletion{T}ZEA\deletion{ }DA
 & KORRONTE\deletion{ }ETIKO HORREN \deletion{H}ELBURUA ZORIONTASUNA LORTZEA DA \\ \hline

\makecell[c]{Ndebele \\ (nbl) }  & Normal  & UMNQOPHO WOMSEBENZI LO & UMN\substitution{C}OPHO \insertion{O}WOMSE\substitution{V}EN\insertion{D}ZI LO\insertion{U}
 & UMN\substitution{C}OPHO WOMSEBENZI LO \\ \hline
 
  \end{tabularx}
  \caption{The ground truth, predictions from XLSR and CA-SSLR models. Deletions are shown with red strikethrough text, insertions are underlined in blue, and substitutions are marked with yellow highlighting.}
  \label{tab:asr_text}
\end{table*}

\label{sec:appendix}

\end{document}